\documentclass[twoside,twocolumn,9pt]{article}
\usepackage{extsizes}
\usepackage[super,sort&compress,comma]{natbib} 
\usepackage[version=3]{mhchem}
\usepackage[left=1.5cm, right=1.5cm, top=1.785cm, bottom=2.0cm]{geometry}
\usepackage{balance}
\usepackage{mathptmx}
\usepackage{sectsty}
\usepackage{graphicx} 
\usepackage{lastpage}
\usepackage[format=plain,justification=justified,singlelinecheck=false,font={stretch=1.125,small,sf},labelfont=bf,labelsep=space]{caption}
\usepackage{float}
\usepackage{fancyhdr}
\usepackage{fnpos}
\usepackage[english]{babel}
\addto{\captionsenglish}{%
  
}
\usepackage{array}
\usepackage{droidsans}
\usepackage{charter}
\usepackage[T1]{fontenc}
\usepackage[usenames,dvipsnames]{xcolor}
\usepackage{setspace}
\usepackage[compact]{titlesec}
\usepackage{hyperref}
\usepackage{bm}
\usepackage{amssymb,amsfonts,amsmath}
\usepackage{subfigure}
\usepackage{stmaryrd}
\allowdisplaybreaks
\usepackage{epstopdf}%This line makes .eps figures into .pdf - please comment out if not required.

\definecolor{cream}{RGB}{222,217,201}

\begin{document}

%\pagestyle{fancy}
%\thispagestyle{plain}
%\fancypagestyle{plain}{
%%%HEADER%%%
%\renewcommand{\headrulewidth}{0pt}
%}
%%%END OF HEADER%%%

%%%PAGE SETUP - Please do not change any commands within this section%%%
\makeFNbottom
\makeatletter
\renewcommand\LARGE{\@setfontsize\LARGE{15pt}{17}}
\renewcommand\Large{\@setfontsize\Large{12pt}{14}}
\renewcommand\large{\@setfontsize\large{10pt}{12}}
\renewcommand\footnotesize{\@setfontsize\footnotesize{7pt}{10}}
\makeatother

\renewcommand{\thefootnote}{\fnsymbol{footnote}}
\renewcommand\footnoterule{\vspace*{1pt}% 
\color{cream}\hrule width 3.5in height 0.4pt \color{black}\vspace*{5pt}} 
\setcounter{secnumdepth}{5}

\makeatletter 
\renewcommand\@biblabel[1]{#1}            
\renewcommand\@makefntext[1]% 
{\noindent\makebox[0pt][r]{\@thefnmark\,}#1}
\makeatother 
\renewcommand{\figurename}{\small{Fig.}~}
\sectionfont{\sffamily\Large}
\subsectionfont{\normalsize}
\subsubsectionfont{\bf}
\setstretch{1.125} %In particular, please do not alter this line.
\setlength{\skip\footins}{0.8cm}
\setlength{\footnotesep}{0.25cm}
\setlength{\jot}{10pt}
\titlespacing*{\section}{0pt}{4pt}{4pt}
\titlespacing*{\subsection}{0pt}{15pt}{1pt}
%%%END OF PAGE SETUP%%%

%%%FOOTER%%%
%\fancyfoot{}
%\fancyfoot[LO,RE]{\vspace{-7.1pt}\includegraphics[height=9pt]{head_foot/LF}}
%\fancyfoot[CO]{\vspace{-7.1pt}\hspace{13.2cm}\includegraphics{head_foot/RF}}
%\fancyfoot[CE]{\vspace{-7.2pt}\hspace{-14.2cm}\includegraphics{head_foot/RF}}
%\fancyfoot[RO]{\footnotesize{\sffamily{1--\pageref{LastPage} ~\textbar  \hspace{2pt}\thepage}}}
%\fancyfoot[LE]{\footnotesize{\sffamily{\thepage~\textbar\hspace{3.45cm} 1--\pageref{LastPage}}}}
%\fancyhead{}
%\renewcommand{\headrulewidth}{0pt} 
%\renewcommand{\footrulewidth}{0pt}
%\setlength{\arrayrulewidth}{1pt}
%\setlength{\columnsep}{6.5mm}
%\setlength\bibsep{1pt}
%%%END OF FOOTER%%%

%%%FIGURE SETUP - please do not change any commands within this section%%%
\makeatletter 
\newlength{\figrulesep} 
\setlength{\figrulesep}{0.5\textfloatsep} 

\newcommand{\topfigrule}{\vspace*{-1pt}% 
\noindent{\color{cream}\rule[-\figrulesep]{\columnwidth}{1.5pt}} }

\newcommand{\botfigrule}{\vspace*{-2pt}% 
\noindent{\color{cream}\rule[\figrulesep]{\columnwidth}{1.5pt}} }

\newcommand{\dblfigrule}{\vspace*{-1pt}% 
\noindent{\color{cream}\rule[-\figrulesep]{\textwidth}{1.5pt}} }

\makeatother
%%%END OF FIGURE SETUP%%%

%%%TITLE, AUTHORS AND ABSTRACT%%%
\twocolumn[
  \begin{@twocolumnfalse}
%{\includegraphics[height=30pt]{head_foot/SM}\hfill\raisebox{0pt}[0pt][0pt]{\includegraphics[height=55pt]{head_foot/RSC_LOGO_CMYK}}\\[1ex]
%\includegraphics[width=18.5cm]{head_foot/header_bar}}\par
\vspace{1em}
\sffamily
%\begin{tabular}{m{4.5cm} p{13.5cm} }

\noindent\LARGE{\textbf{The interplay between symmetry-breaking and symmetry-preserving bifurcations in soft dielectric films and the emergence of giant electro-actuation}} \\%Article title goes here instead of the text "This is the title"

To appear in Extreme Mechanics Letters (\href{https://doi.org/10.1016/j.eml.2020.101151}{DOI:10.1016/j.eml.2020.101151})\\

%\vspace{0.3cm}  \\

\noindent\large{Lingling Chen$^1$, Xu Yang$^1$, Binglei Wang$^1$, Shengyou Yang$^{1{,}2}$$^{\ast}$, Kaushik Dayal$^{3{,}4{,}5}$, Pradeep Sharma$^{6{,}7}$$^\dagger$ }\\%Author names go here instead of "Full name", etc.

\noindent\normalsize{ Soft elastomers that can exhibit extremely large deformations under the action of an electric field are essential for applications such as soft robotics, stretchable and flexible electronics, energy harvesting among others. The critical limiting factor in conventional electro-actuation of such materials is the occurrence of the so-called pull-in instability. In this work, we demonstrate an extraordinarily simple way to coax a dielectric thin film towards a symmetry-breaking pitchfork bifurcation state while avoiding pull-in instability. Through the nonlinear interplay between the two bifurcation modes, we predict electro-actuation strains that exceed what is conventionally possible by 200$\%$, and at significantly lower applied electric fields.} \\%The abstrast goes here instead of the text "The abstract should be..."

%\end{tabular}

 \end{@twocolumnfalse} \vspace{0.6cm}

  ]
%%%END OF TITLE, AUTHORS AND ABSTRACT%%%

%%%FONT SETUP - please do not change any commands within this section
\renewcommand*\rmdefault{bch}\normalfont\upshape
\rmfamily
\section*{}
\vspace{-1cm}

%%%FOOTNOTES%%%

\footnotetext{\textit{$^{1}$~~Department of Engineering Mechanics, School of Civil Engineering, Shandong University, Jinan, 250061, China\\
$^{2}$~~Suzhou Research Institute, Shandong University, Jiangsu, 215123, China\\
$^{3}$~~Center for Nonlinear Analysis, Department of Mathematical Sciences, Carnegie Mellon University, PA, USA\\
$^{4}$~~Department of Materials Science and Engineering, Carnegie Mellon University, PA, USA\\
$^{5}$~~Department of Civil and Environmental Engineering, Carnegie Mellon University, PA, USA\\
$^{6}$~~Department of Mechanical Engineering, University of Houston, TX, USA\\
$^{7}$~~Department of Physics, University of Houston, TX, USA\\
$^{\ast}$syang\_mechanics@sdu.edu.cn (corresponding author)\\
$^\dag$psharma@uh.edu (corresponding author)}
}

%Please use \dag to cite the ESI in the main text of the article.
%If you article does not have ESI please remove the the \dag symbol from the title and the footnotetext below.
%\footnotetext{\dag~Electronic Supplementary Information (ESI) available: [details of any supplementary information available should be included here]. See DOI: 10.1039/cXsm00000x/}
%additional addresses can be cited as above using the lower-case letters, c, d, e... If all authors are from the same address, no letter is required

%\footnotetext{\ddag~Additional footnotes to the title and authors can be included \textit{e.g.}\ `Present address:' or `These authors contributed equally to this work' as above using the symbols: \ddag, \textsection, and \P. Please place the appropriate symbol next to the author's name and include a \texttt{\textbackslash footnotetext} entry in the the correct place in the list.}

%%%END OF FOOTNOTES%%%

%%%MAIN TEXT%%%%
%%%%%%%%%%%%%%%
\section{Introduction}
Soft materials such as elastomers have elastic moduli that can be several orders of magnitude smaller than conventional polymers. 
Consequently, materials like silicone are able to easily sustain large deformations under the action of relatively modest forces. 
For several applications e.g. soft robotics \cite{rus2015design, shian2015dielectric} or actuators \cite{pelrine2000high, pelrine2000highfield, kofod2007energy, keplinger2010rontgen, brochu2010advances}, we require the deformation to take place in response to an electric field. 
In a well-known study, Keplinger {\em et al}. \cite{keplinger2012harnessing} obtained an areal increase of almost 1700 $\%$  for a specially fabricated acrylic membrane under the action of a suitably high electric field. 
Over the past two decades, research on electro-active dielectric elastomers has significantly intensified due to potential applications like energy harvesting \cite{koh2009maximal, bauer201425th, deng2014nanoscale, yang2017avoiding, invernizzi2016energy}, adaptive optics \cite{carpi2010stretching}, and stretchable electronics \cite{rogers2010materials, lu2014flexible}.\\

\begin{figure}[t] % 
    \centering
    \includegraphics[width= 1 \linewidth]{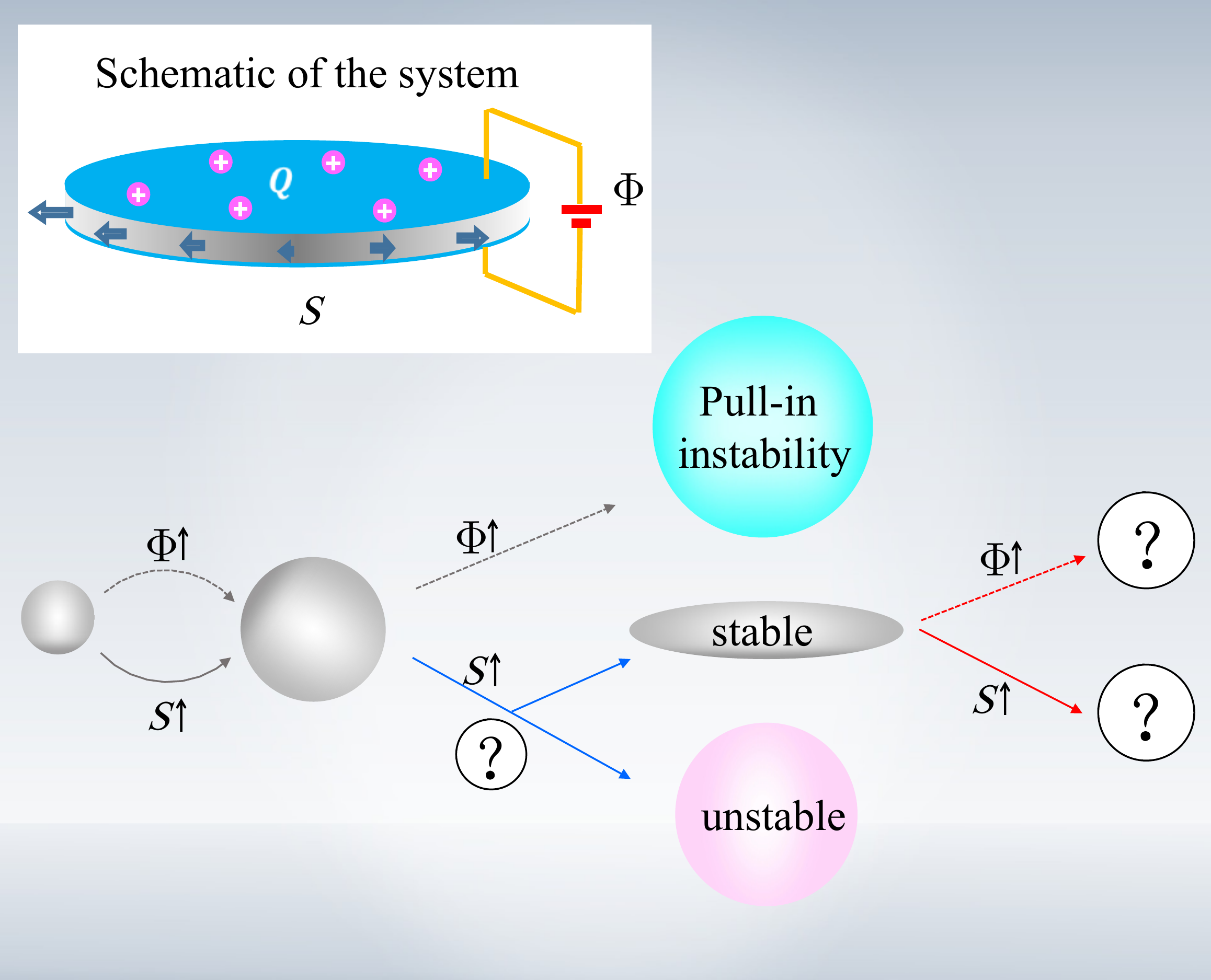}
    \caption{\small Schematic of the central idea: We consider the various deformation paths that a circular thin film dielectric elastomer can take under the combined action of an applied voltage $\Phi$ in the thickness direction and an in-plane symmetric radial dead load $S$. 
    A gradually increasing voltage ($\Phi$$\uparrow$) will lead to pull-in instability, however, a gradually increasing dead load ($S$$\uparrow$) will lead to the so-called Treloar-Kearsley (T-K) instability at which the stable (circular) film will bifurcate into a stable \emph{elliptical} configuration.}
\label{fig1}
\end{figure}

With the large deformations and highly nonlinear behavior characteristic of elastomers, we must also contend with the inevitability of instabilities such as surface wrinkles \cite{biot1963surface, wang2013creasing, godaba2017dynamic}, creases \cite{hohlfeld2011unfolding, wang2011creasing, zurlo2017catastrophic}, folds \cite{pocivavsek2008stress, kim2011hierarchical}, electro-buckling \cite{tavakol2016voltage, yang2017revisiting}, pull-in instability, bursting drops in solid dielectrics \cite{wang2012bursting}, and so on. 
We specifically highlight the {\it pull-in instability} \cite{stark1955electric, plante2006large, zhao2007method, xu2010electromechanical, li2011electromechanical, huang2012giant, jiang2016eliminating} due to its singular role as a  ``failure mode" in dielectric elastomers and the key limiting factor that limits electro-actuation and energy harvesting. \\

Consider a thin dielectric film subjected to a fixed potential difference across its thickness. 
The applied field will compress the film -- due to Maxwell stress, or more generally electrostriction -- thereby decreasing the film thickness and causing a lateral expansion. 
The thinning of the film increases the magnitude of the electric field in the material, in turn leading to a further decrease in thickness.
When the film thickness decreases to a critical value, this coupling leads to a pull-in instability.
At this critical value, the thinning increases dramatically, and the magnitude of the electric field causes electrical breakdown.
Mathematically, the pull-in instability can be regarded as a {\it limit point or saddle-node bifurcation} \cite{Nicolis1995Introduction}.
That is, the stable and unstable branches ``collide'' at the critical electric field, and subsequently are annihilated as the electric field increases beyond the critical value. \\

Traditionally, researchers have devised strategies to \emph{suppress} pull-in instability to improve electro-actuation, e.g., by using pre-stretch; among many papers, we highlight recent approaches including \cite{zhao2014harnessing, dorfmann2017nonlinear, dorfmann2019instabilities, LU2020100752}. 
In this work, we break from the conventional philosophy of suppressing pull-in instabilities, and instead seek to \emph{exploit} an interplay between pull-in instability and another instability described below.
We propose to guide the soft material system towards an alternate symmetry-breaking instability -- a super-critical pitchfork bifurcation -- to obtain an unprecedented level of electro-actuation not possible with current approaches. \\

The key concept that we propose has its roots in a purely mechanical experiment by Treloar about half a century ago \cite{Treloar1949stresses}. 
He studied a square rubber sheet stretched (within its plane) by equal forces on the lateral faces, corresponding to dead loading.
Treolar, however, observed that upon reaching a critical load, the square sheet deformed into a symmetry-breaking rectangular sheet.
Of course, {\em linear} elasticity predicts that the square sheet remains square regardless of the magnitude of the equibiaxial forces.
This surprising experimental observation motivated a number of follow-up theoretical and experimental works, c.f. \cite{Kearsley1986asymm, chen1987stability, batra2005treloar, Steigmann2007simple}.
Kearsley \cite{Kearsley1986asymm}, using nonlinear elasticity, showed theoretically that indeed, beyond a critical load, the elastic sheet admits a stable symmetry-breaking deformed state and the symmetric response becomes unstable. 
This phenomenon -- the {\it Treloar-Kearsley (T-K) instability} -- is a {\it supercritical pitchfork bifurcation} \cite{strogatz2014nonlinear}.
To date, this instability has not be investigated in the context of coupled electro-mechanical behavior of soft materials. \\

Figure \ref{fig1} illustrates our key concept. 
Consider a dielectric elastomer thin film. 
For both illustration and subsequent calculations, we will assume the film to be initially circular. 
If we increase the applied voltage ($\Phi$$\uparrow$) slightly, the circular film will become thinner and expand laterally into a larger circular film. 
If we continue to increase the voltage ($\Phi$$\uparrow$), the larger circular film will eventually undergo a pull-in instability (at some critical voltage that depends on the material parameters and the film thickness). 
However, we may, instead of a voltage difference, apply a radially-symmetric in-plane dead-load to the thin film, analogous to the experiment by Treloar. 
Increasing the magnitude of the dead load ($S$$\uparrow$) will also deform the circular film to a larger circular film until the occurrence of the T-K instability. 
At that point, the circular configuration will be unstable and the film will deform into an ellipse. \\

The loading configurations depicted in Figure \ref{fig1} and described in the preceding paragraph naturally bring up the following questions, in terms of the various combinations of electromechanical loads and possible instabilities:
(1) Which instability (pull-in or T-K) will occur first and how does the  constitutive law of the material affect their occurrence?
(2) If we continue to increase the electromechanical load after the onset of T-K instability, what are the equilibrium states? 
(3) For a given electromechanical load, what are the equilibrium solutions, and which are stable?
(4) Most importantly, can we exploit the interplay between the two instabilities to obtain large electro-actuation? 
In the remainder of this paper, we will quantitatively address these questions. \\

%%%%%%%%%%%%%%%
\section{Formulation} \label{sec-formulation}

We briefly summarize the key points of our formulation of the central problem with details being relegated to Appendix A. We remark that there have been intriguing recent developments \cite{grasinger2020statistical, grasinger2020architected} that address the electromechanical coupling of dielectric elastomers starting from the monomer level and then upscale to a coarse-grained description using statistical mechanics. In this work, we follow a purely phenomenological macroscopic description of the elastomers. To that end, consider a circular dielectric film with radius $R$ and thickness $H$ in the reference undeformed state, subject to both an applied potential difference $\Phi$ across its thickness and in-plane mechanical tractions.
We denote the reference and deformed coordinates by $\bm X$ and $\bm x$, and define the deformation gradient $\bm F = \partial \bm x / \partial \bm X$.
We restrict our attention to homogeneous deformations with $\bm F$ constant, thereby automatically satisfying force equilibrium if we satisfy the boundary conditions.
We use Cartesian coordinates $(X_1, X_2, X_3)$ with an orthonormal basis $({\bm e}_1, {\bm e}_2, {\bm e}_3)$ that are aligned along the principal directions of $\bm F^T \bm F$, and ${\bm e}_3$ is the direction of the film normal.
The deformation then has the simple form $\bm x = (\alpha_1 X_1, \alpha_2 X_2, \alpha_3 X_3)$, where $\alpha_i >0$, $i =1,2,3$, are the (constant) principal stretches, and we assume without loss of generality that $\alpha_1 \ge \alpha_2$.
Further, assuming incompressibility -- a very good approximation for elastomers -- requires that the Jacobian $J = \det {\bm F} = 1$, i.e., $ \alpha_1 \alpha_2 \alpha_3 = 1$.
We also define the domain of the circular film in the reference configuration as $\mathcal B_0 = \{\bm X: 0 \le \sqrt{X_1^2 + X_2^2} \le R, \ 0 \le X_3 \le H \}$. \\

The total nominal stress ${\bm T}$ within the incompressible dielectric elastomers is then \cite{suo2008nonlinear, zhao2014harnessing}:
\begin{equation} 
\label{eqn:sy-10}
    {\bm T} = \frac{\partial W^e}{\partial {\bm F}} + {\bm T}^M - \kappa {\bm F}^{-T}.
\end{equation}
Here $W^e$ is the purely mechanical contribution to the energy density, ${\bm T}^M$ is the nominal electric Maxwell stress, $\kappa$ is the Lagrange multiplier conjugate to the incompressibility constraint, and ${\bm F}^{-T}$ is the inverse of the transpose of ${\bm F}$. 
The nominal Maxwell stress is given by ${\bm T}^M := \frac{1}{2} \varepsilon {\tilde E}^2 \alpha_3^{-2} {\rm diag} \,  (-\alpha_1^{-1}, -\alpha_2^{-1}, \alpha_3^{-1} )$, where $\varepsilon$ is the dielectric permittivity and $\tilde E = \Phi/H$. \\

Assuming isotropy, $W^e$ in \eqref{eqn:sy-10} depends on $\bm F$ only through the principal stretches $\alpha_1$, $\alpha_2$ and $\alpha_3$. 
Using the fact that the top and bottom faces are traction-free, i.e. $\bm T {\bm e}_3 = {\bm 0}$, we can readily solve for the Lagrange multiplier $\kappa = \alpha_3 {\partial W^e}/{\partial \alpha_3} + {\varepsilon {\tilde E}^2 \alpha_3^{-2}}/{2}$. 
%Since the deformation and stresses are independent of position, force equilibrium ${\rm Div} \, \bm T = \bm 0$ is automatically satisfied. 
The given radial traction with magnitude $S$ applied on the lateral faces, together with the constraint of incompressibility, forms a system of algebraic equations with unknowns $(\alpha_1, \alpha_2, \alpha_3)$ for a given applied stimulus $(\tilde E, S)$. \\

The material constitutive law is implicit in the specification of the strain-energy function. In what follows, we will consider a Mooney-Rivlin solid \cite{rivlin1948large} ($W^e = \frac{\mu}{2} \sum_{i=1}^3 \left\{ \left(\alpha_i^2 -1 \right) + \gamma \left(\alpha_i^{-2} -1 \right) \right\}$) which also subsumes the often used neo-Hookean model. Here $\mu$ and $\gamma$ are the elastic material constants. The case of $\gamma =0$ corresponds to a neo-Hookean solid.  We refer the reader to Appendix B regarding the choice of the constitutive law for the problem at hand.\\

Since both the mechanical loads and the applied electrical field are consistent with radial symmetry in the plane, we might anticipate that the equilibrium configuration will also be symmetric, i.e., $\alpha_1 =\alpha_2$. 
However, to allow for the possibility of symmetry-breaking deformations, we allow $\alpha_1\neq\alpha_2$, and find the following relation using the equilibrium condition (see Appendix A for details):
\begin{equation} \label{sugg}
    \begin{aligned}
    0 &= (\alpha_1 - \alpha_2) \times \Big[1 + \Big( \frac{\varepsilon {\tilde E}^2}{\mu} -\gamma \Big) \alpha_1 \alpha_2 \\
    & \quad + \alpha_1^{-3}\alpha_2^{-3} \left[ 1+  \gamma \left( \alpha_1^{2} + \alpha_1 \alpha_2 + \alpha_2^{2} \right) \right] \Big].
    \end{aligned}
\end{equation}\\

The key point is that the second term on the right of \eqref{sugg} can be zero under some conditions, and hence allows for $\alpha_1\neq\alpha_2$.
For the case of neo-Hookean solids ($\gamma=0$), it is evident that the second term on the right of \eqref{sugg} is always positive, which then mandates $\alpha_1 =\alpha_2$.
In contrast, the more general Mooney-Rivlin solid can admit both symmetric and symmetry-breaking equilibrium solution.\\

%%%%%%%%%%%%%%%%%%%%%%%%%%%%%
\begin{figure}[ht!] % 
    \centering
    \subfigure[]{%
    \includegraphics[width=0.85 \linewidth]{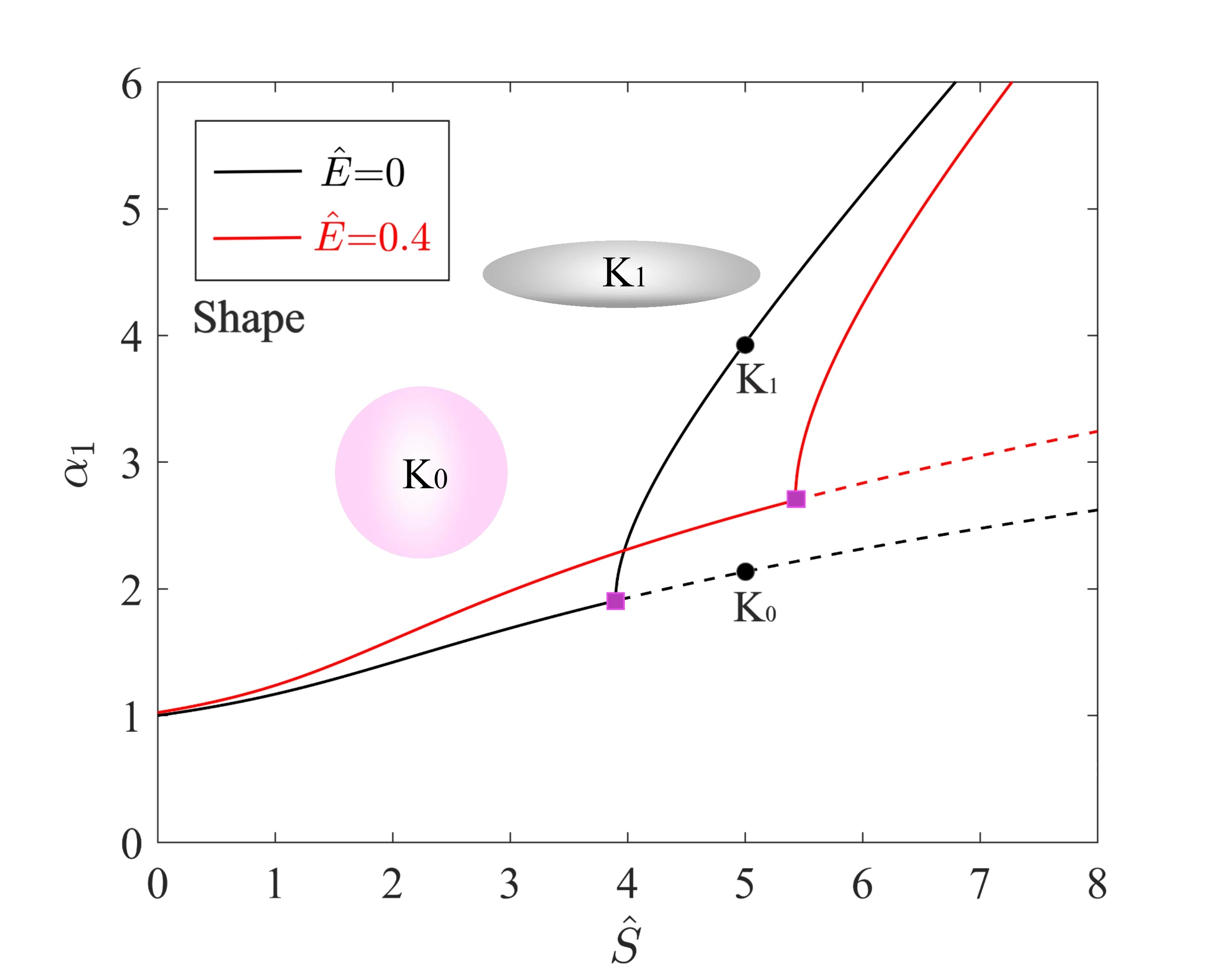} \label{fig2a}}
    \subfigure[]{%
    \includegraphics[width=0.85 \linewidth]{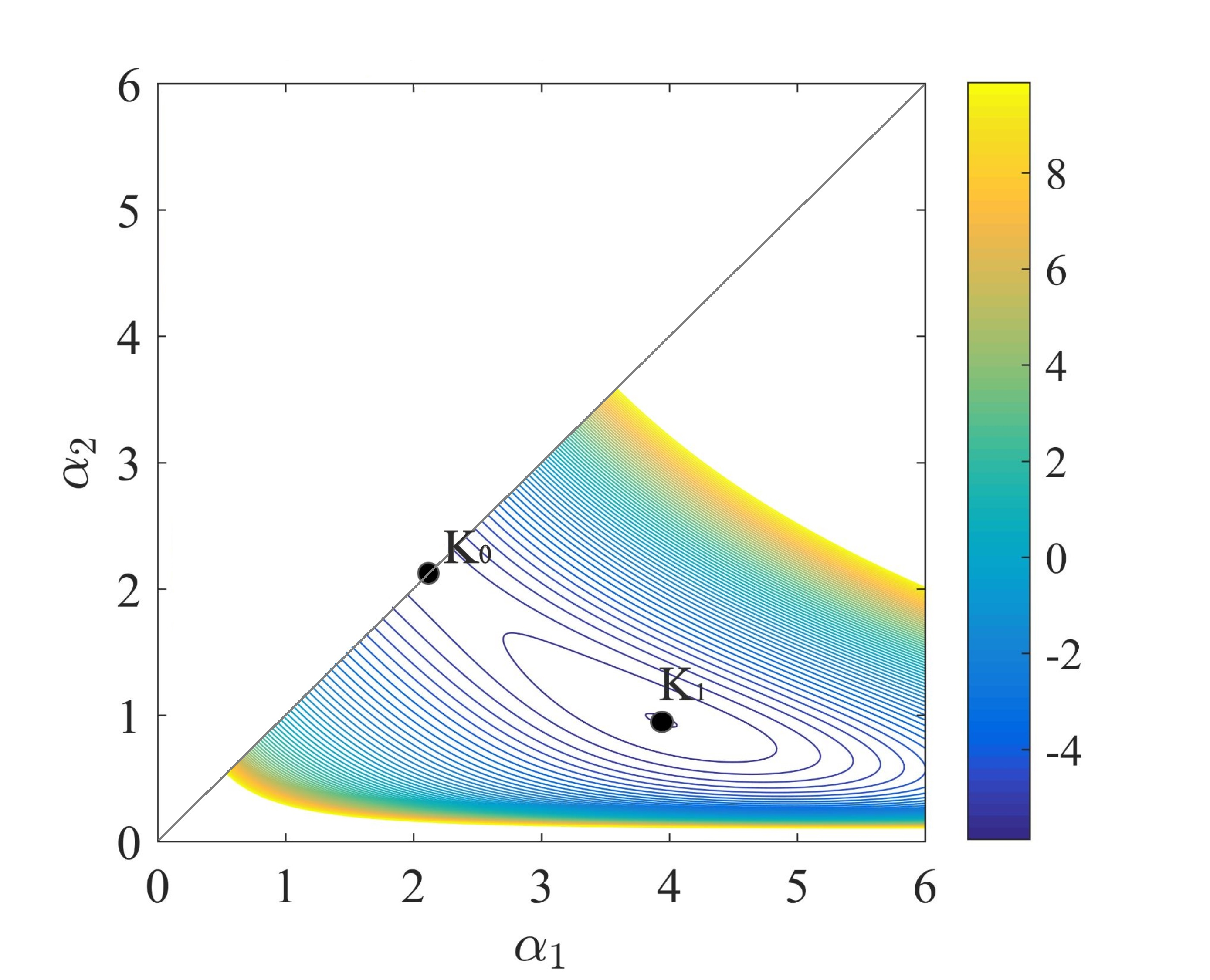} \label{fig2b}}
    \caption{\small The bifurcation diagram. (a) The stretch $\alpha_1$ as a function of the dead load $\hat S$. The solid curves are the stable equilibrium states, corresponding to either a circular film ($\alpha_1 = \alpha_2$) or an elliptical film ($\alpha_1 \neq \alpha_2$). The dashed curves are the unstable equilibrium states, corresponding to the T-K instability. The critical point for the onset of T-K instability is marked by $\square$. (b) The contours of the total energy in the $\alpha_1 - \alpha_2$ plane at $(\hat S, \hat E) = (5, 0)$. The point labeled K$_1$ is the stable asymmetric elliptical configuration, and K$_0$ is the unstable symmetric circular configuration.}
\label{fig2}
\end{figure}

%%%%%%%%%%%%%%%%%
%%%%%%%%%%%%%%%%%Appendix
\section{Linear bifurcation analysis} \label{sec-LinearBF}
%Before we proceed to perform numerical calculations to illustrate our central idea and address the questions raised earlier, it is instructive first to perform a linear bifurcation analysis.  

We next use a linear bifurcation analysis, before turning to a numerical approach further below.
While the linear bifurcation analysis can only give us the necessary conditions for the onset of bifurcation and will not allow us to distinguish between pull-in instability (a limit point bifurcation) and T-K instability (a pitchfork bifurcation), the closed-form calculations provide physical insights and also guide the numerical calculations.\\

In the linearized setting, the equilibrium equations reduce to:
\begin{equation} \label{eq}
    L_j (\alpha_1, \alpha_2; \tilde E, S) = T_{j} (\alpha_1, \alpha_2; \tilde E)  - S = 0,
\end{equation}
where $j=1,2$ and $T_{j}$ are the principal stresses. 
We can determine the onset of bifurcation by examining the conditions for $\det(\frac{\partial L_i}{\partial \alpha_j})=0$. 
At the onset of bifurcation, we have that $\alpha_1 = \alpha_2 = \alpha$ since we are linearizing about the symmetric state.
The condition for the onset of bifurcation then becomes (see Appendix C): 
\begin{equation} \label{condition}
\begin{aligned}
    0 & = \left( 1 +3 \gamma \alpha^{-4} +5 \alpha^{-6} + 3(\gamma - \frac{\varepsilon {\tilde E}^2}{\mu}) \alpha^{2} \right) \\
    & \quad \times \left( 1 +3 \gamma \alpha^{-4} + \alpha^{-6} - (\gamma - \frac{\varepsilon {\tilde E}^2}{\mu}) \alpha^{2} \right).
\end{aligned}
\end{equation}\\

%%%%%%%%%%%%%%%%%
%%%%%%%%%%%%%%%%%Appendix
%\subsection{Pull-in instability or a pitchfork bifurcation} \label{appendix-pullin-vs-asymmetric}
Using that $\gamma \ge 0$ for real materials, we can distinguish three cases:
\begin{enumerate}
    \item 
    When $\frac{\varepsilon {\tilde E}^2}{\mu} = \gamma$, the equality \eqref{condition} does not hold for any $\alpha >0$ since both the first and second terms in \eqref{condition} are positive. In other words, this case implies the nonexistence of bifurcations, and neither the T-K instability nor pull-in instability will occur.

    \item 
    When $\frac{\varepsilon {\tilde E}^2}{\mu} > \gamma$, the second term on the right of \eqref{condition} is always positive while the first term may be zero under some conditions. Setting the first term on the right of \eqref{condition} to zero and using also the equilibrium equations, we can determine the threshold and the corresponding value of $\alpha>1$ at which the bifurcation occurs. Since the second term on the right of \eqref{sugg} is positive, this case only admits the solution $\alpha_1 =\alpha_2$ and the bifurcation point corresponds to a limit point, i.e., the onset of pull-in instability.

    \item 
    When $\frac{\varepsilon {\tilde E}^2}{\mu} < \gamma$, the first term on the RHS of \eqref{condition} is always positive while the second term can be zero under some conditions. The zero second term on the RHS of \eqref{condition}, together with the equilibrium condition, we can determine the threshold and the corresponding value of $\alpha>1$ at which the bifurcation occurs. Since the second term on the RHS of \eqref{sugg} and the second term on the RHS of \eqref{condition} become zero simultaneously, the bifurcation point in this case corresponds to the onset of Treloar-Kearsley instability. 
\end{enumerate}

%%%%%%%%%%%%%%%
\section{Results and discussions} \label{sec-RD}
We now turn to numerical calculations to make further progress (see Appendix D for the details of the approach).
In what follows, we choose the material constant of $\gamma=0.3$ for all the numerical plots, but this particular choice does not impact the central conclusions drawn in this paper. The nondimensional dead load and electric field are defined by $\hat S = \frac{S}{\mu}$ and $\hat E = \frac{\tilde E}{\sqrt{\mu/\varepsilon}}$, respectively.\\

Figure \ref{fig2a} shows the bifurcation diagram. 
At zero electric field and dead load $\hat S$ in the range of $(0, 3.898)$, only the symmetric state ($\alpha_1=\alpha_2$) is a stable equilibrium (see the appendices for more details). 
The equal stretches both increase monotonically from $1$ to $1.905$ as $\hat S$ increases from $0$ to $3.898$. 
Once $\hat S$ exceeds the threshold of $3.898$, the circular film bifurcates to an elliptical configuration with $\alpha_1 > \alpha_2$.\\
% may bifurcate to either a horizontal elliptical film ($\alpha_1 > \alpha_2$, with semi-major axis in the horizontal direction) or a vertical elliptical film ($\alpha_1 < \alpha_2$, with semi-major axis in the vertical direction). In other words, the circular configuration is unstable, and two new stable elliptical configuration have emerged.

We take the case of $(\hat S, \hat E)=(5, 0)$ as an illustration. There exist two equilibrium states that correspond to two points K$_0$ and K$_1$ in Figure \ref{fig2a}, and the energy contour is shown in Figure \ref{fig2b}. The symmetric deformation with stretches $\alpha_1 =\alpha_2 = 2.135$ at point K$_0$ is an unstable circular film while the symmetry-breaking deformation with stretches $\alpha_1= 3.939$ and $\alpha_2 =0.958$ at point K$_1$ is a stable elliptical film. It is clear that the semi-major axis of the ellipse is about twice the radius of the circle, i.e., $3.939/2.135=1.845$, which indicates a relatively large deformation induced by the instability.\\

In Figure \ref{fig2a}, we also find that the electric field will delay the onset of T-K instability. Without any applied electric field, T-K instability occurs at $\hat S = 3.898$ and the stretch $\alpha_1$ (or $\alpha_2$) is $1.905$. However, under an electric field of $\hat E = 0.4$, T-K instability occurs at $\hat S =5.428$ and the stretch $\alpha_1$ (or $\alpha_2$) is $2.699$; moreover, there exists only one stable circular film with stretches $\alpha_1 = \alpha_2 = 2.591$ at $\hat S = 5$. This implies the nonexistence of bifurcation and the suppression of T-K instability by using an electric field. \\

%%%%%%%%%%%%%%%
%%%%%%%%%%%%%%%
%%%%%%%%%%%%%%%%%%%%%%%%%%%%%
\begin{figure}[ht!] % 
\centering
\subfigure[]{%
\includegraphics[width=0.85 \linewidth]{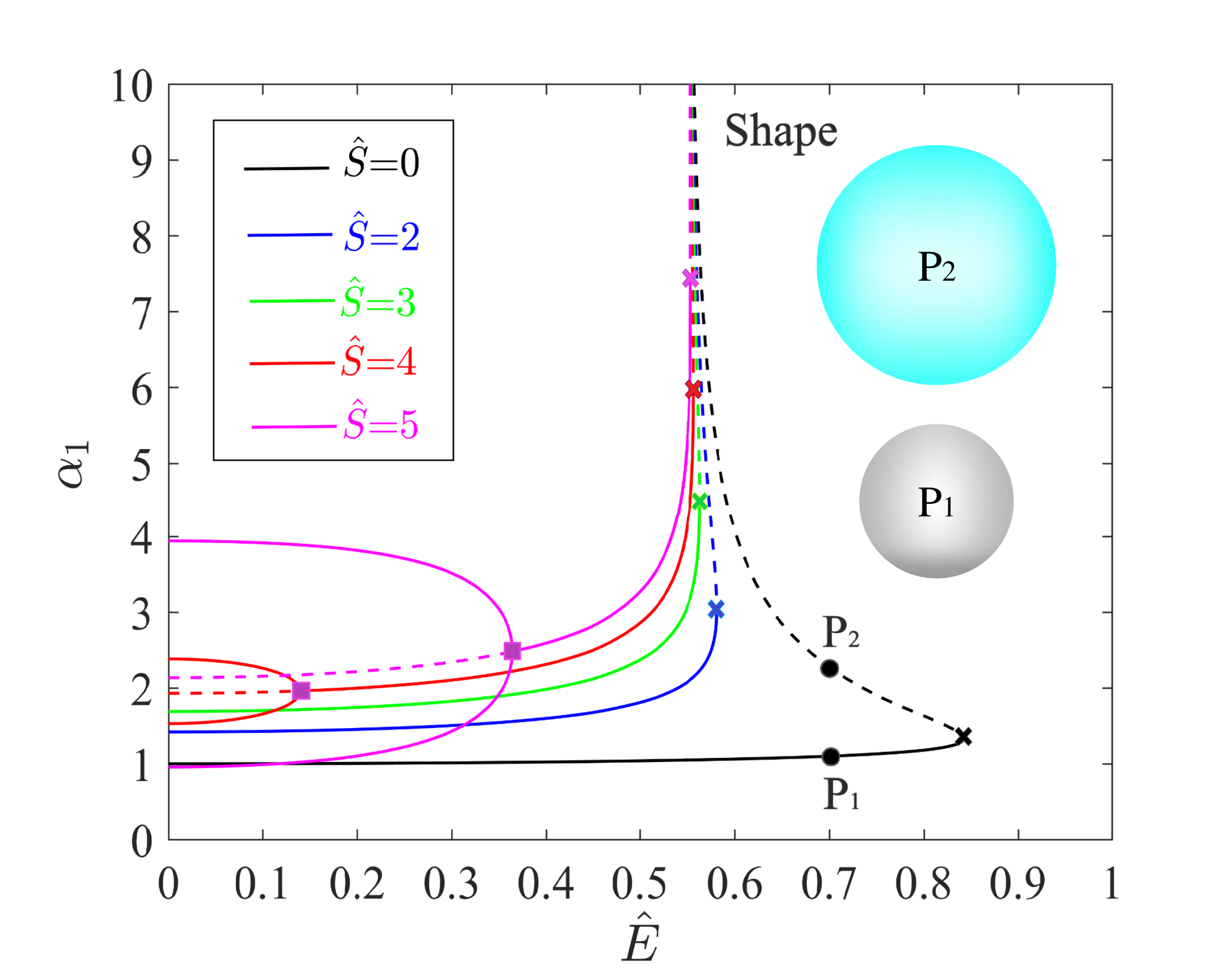} \label{fig3a}}
\\
\subfigure[]{%
\includegraphics[width=0.85 \linewidth]{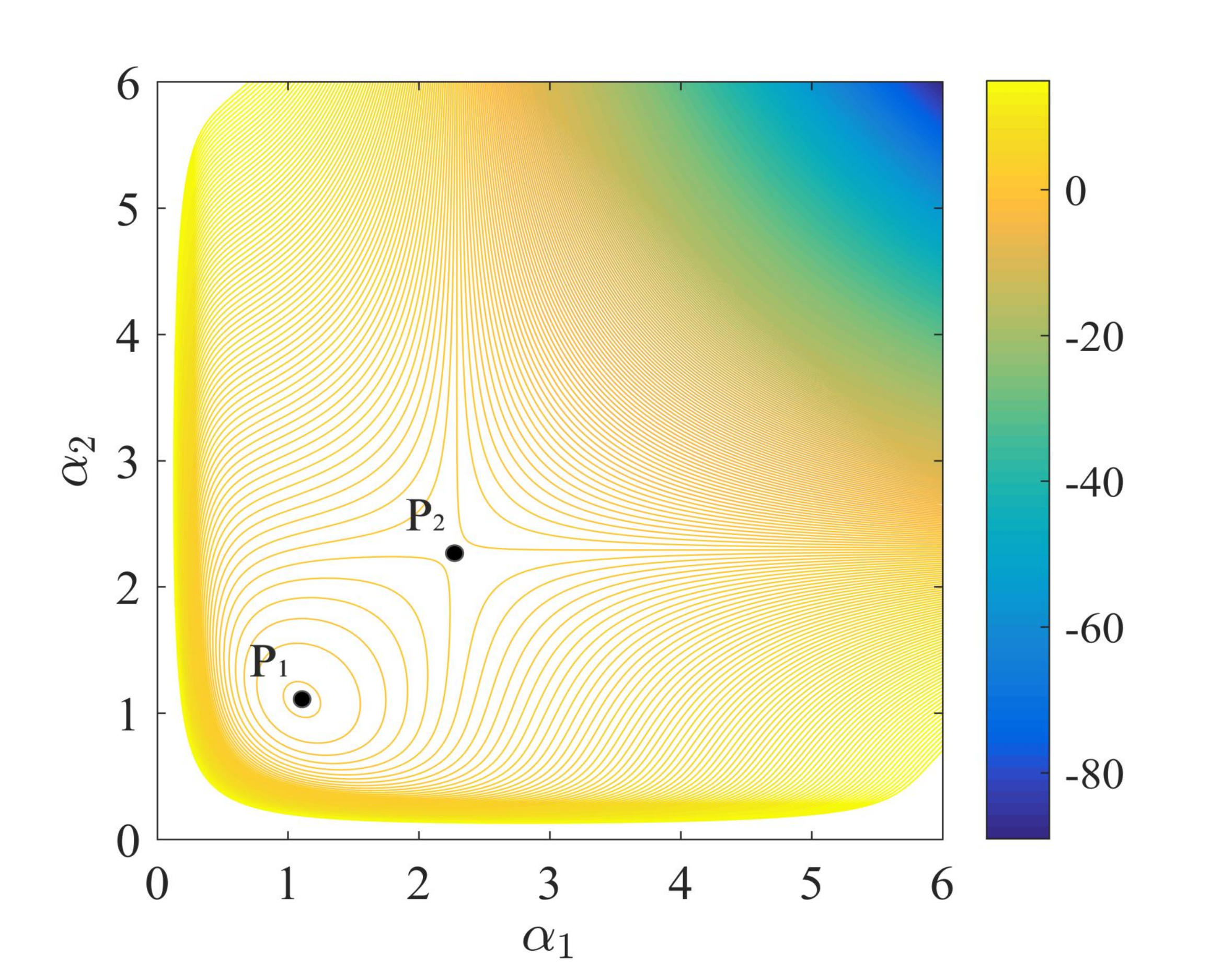} \label{fig3b}}
\caption{\small (a) The nominal electric field $\hat E$ $vs$. the stretch $\alpha_1$. On each curve, the critical point for the onset of pull-in instability is marked by a cross `$\times$'. (b) Contour plot of the total energy of the electrostatic system on the $\alpha_1 - \alpha_2$ plane.}
\label{fig3}
\end{figure}

In Figure \ref{fig3a}, at an applied electric field of $\hat E =0.7$ but a zero dead load, the film has two equilibrium states of symmetric deformations, which are denoted by points P$_1$ and P$_2$. With the energy contour shown in Figure \ref{fig3b}, point P$_1$ is a stable node while point P$_2$ is a saddle point. By increasing the electric field, the two points collide at the threshold of $\hat E = 0.842$ and then pull-in instability occurs. The dead load can assist the electric field in inducing pull-in instability. For example, compared to the curve of $\hat S=0$ with the threshold of $0.842$, the curve of $\hat S=2$ has a critical electric field of $\hat E = 0.580$ for the onset of pull-in instability, which has a much lower threshold. We note that there is no T-K instability in these two curves. \\

In Figure \ref{fig3a}, we also observe that a sufficient large dead load can cause the T-K instability to occur on the $\hat E - \alpha_1$ plane, but it occurs at a lower electric field compared to that of pull-in instability. For the case of $\hat S=0$ (or $2$), we only have pull-in instability that occurs at $\hat E^P =0.842$ (or $0.580$). However, for the case of $\hat S=4$, we can have either T-K instability at $\hat E^{TK} = 0.139$ or pull-in instability at $\hat E^P = 0.556$. For another case of $\hat S=5$, T-K instability occurs at $\hat E^{TK} = 0.365$ while pull-in instability occurs at $\hat E^P = 0.553$. \\

\begin{figure}[ht!] % 
\centering
\subfigure[]{%
\includegraphics[width=0.85 \linewidth]{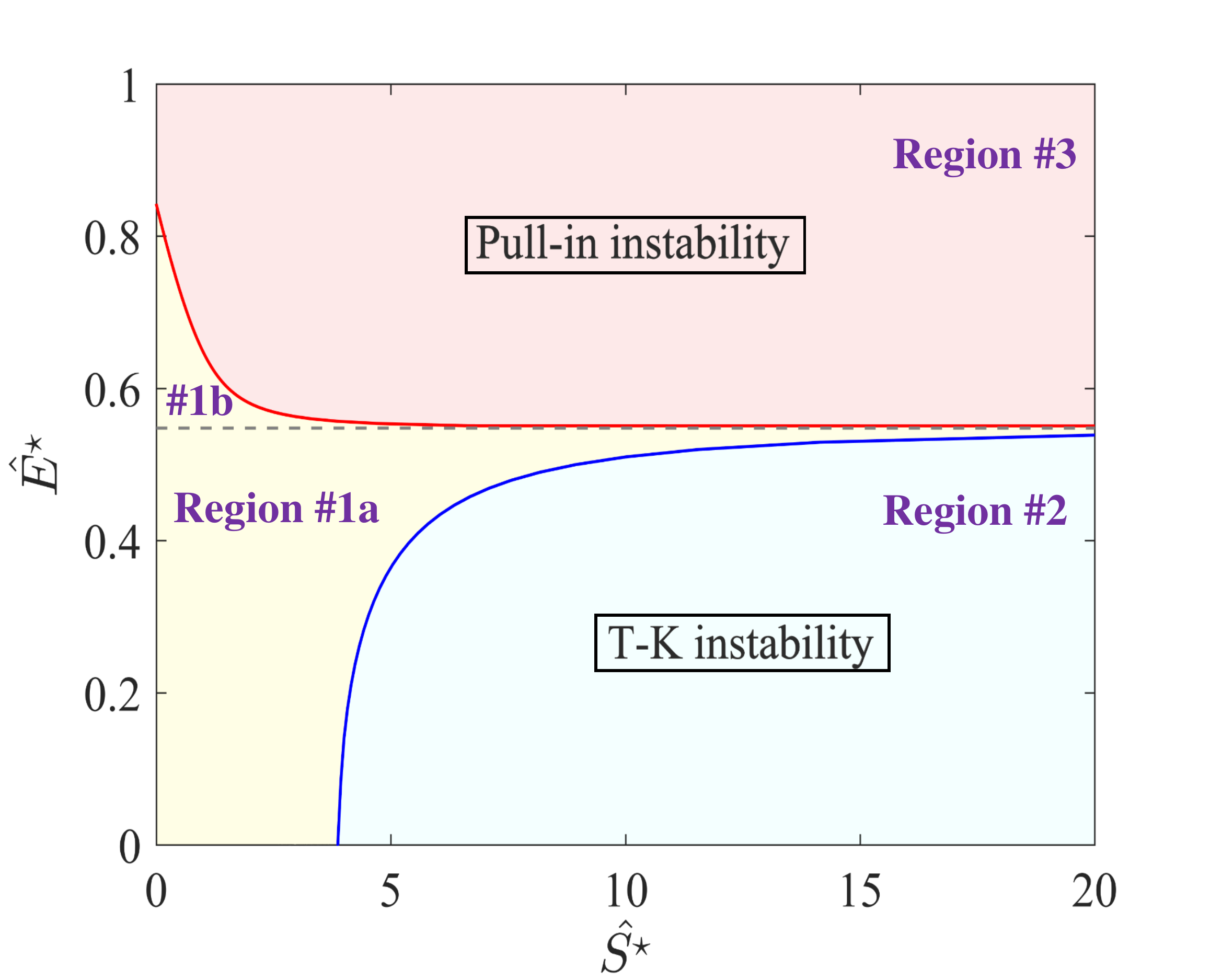} \label{fig4a}}
\subfigure[]{%
\includegraphics[width=0.85 \linewidth]{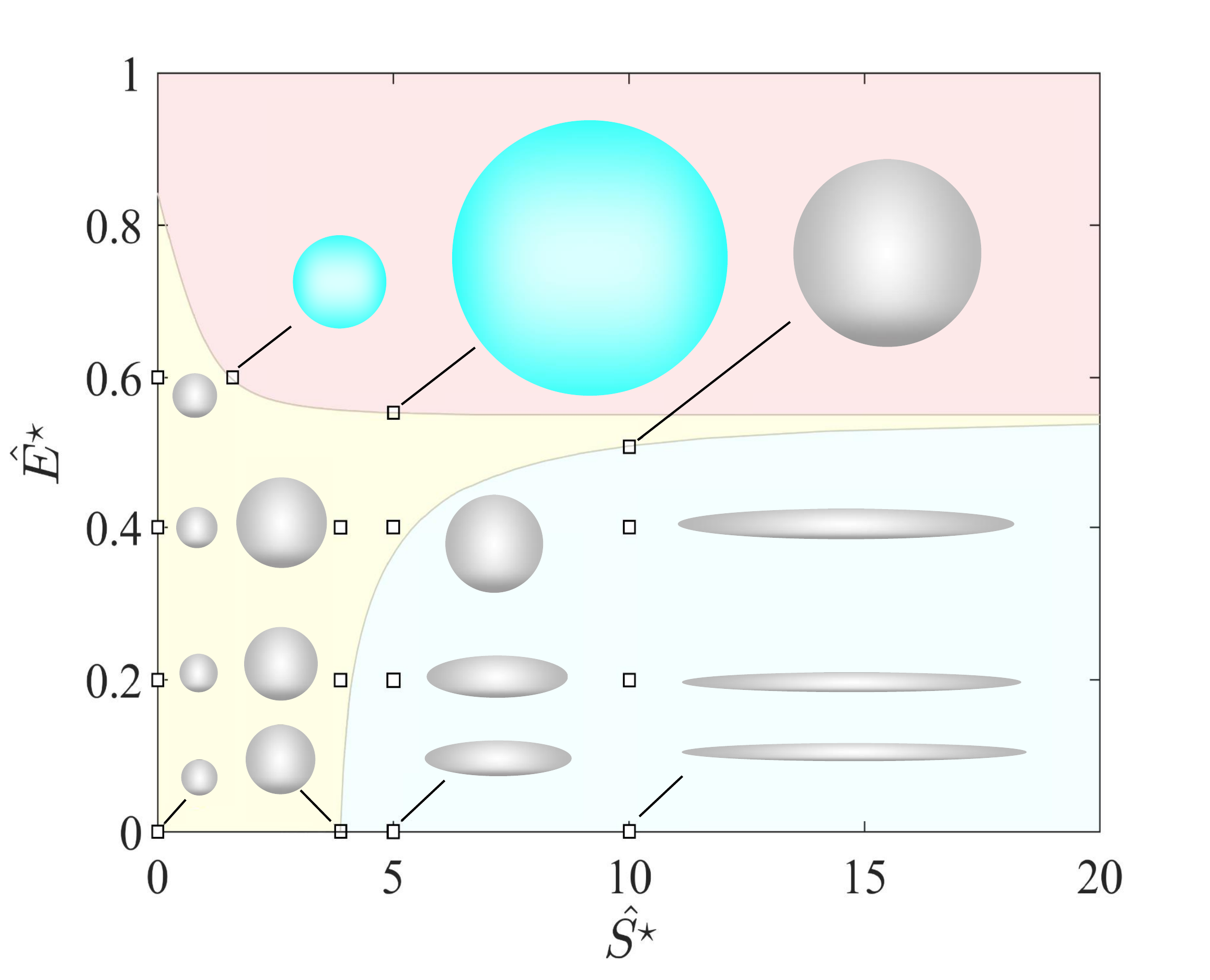} \label{fig4b}}
\caption{\small (a) Phase diagram of a circular film subjected to the mechanical and electric loads. (b) The original circular film at $(\hat S^\star, \hat E^\star) = (0,0)$ is denoted by a solid gray circle. The deformed shapes at some selected points are presented.}
\label{fig4}
\end{figure}

To systematically explore the interplay between T-K instability and pull-in instability, we plot the phase diagram in Figure \ref{fig4a}. The thresholds of T-K instability (blue curve) and pull-in instability (red curve) in Figure \ref{fig4a} can be determined by using linear bifurcation analysis (see the appendices). The two thresholds, i.e., the red and blue curves, separate the load-plane into three regions: region$\#1$, region$\#2$, and region$\#3$. To visualize the deformation process, we select some points on the load-plane and illustrate the corresponding deformations in Figure \ref{fig4b}. Note that the equilibrium states are obtained by solving the two algebraic equations \eqref{T11=S-T22=S-mooney} and their stability is investigated by using the energy method (see the appendices). The number of equilibrium solutions, their stabilities, and the shapes of the equilibrium solutions in each region on the load-plane in Figure \ref{fig4} are summarized as follows: 

\begin{itemize}
\item In region$\#1$, the horizontal line $\hat E^\star = \sqrt{\gamma}$ separates it into two subregions: region$\#1a$ and region$\#1b$. In region$\#1a$, the dielectric film only has {\it one} equilibrium state---a stable circular film. In contrast, the film in region$\#1b$ has {\it two} equilibrium states: the stable state is a smaller circular film and the unstable state is a larger circular film.

\item In region$\#2$, the dielectric film encounters T-K instability and has {\it two} equilibrium states: one unstable state is a circular film and the other state is the elliptical configuration.

\item In region$\#3$, the dielectric film encounters pull-in instability and has no equilibrium solutions.
\end{itemize}

Using the deformation in Figure \ref{fig4b} as a guide, we can now highlight the prospects of giant electro-actuation induced by the electromechanical instability. At a given pair of loads $(\hat S^\star, \hat E^\star) = (5, 0.4)$, the original circular film (with a radius of 1) deforms to a stable circular film with a larger radius of $2.591$. Interestingly, if we fix the dead load but decrease the electric field from $0.4$ to $0.2$, the original film deforms to a stable elliptical film with a semi-major axis of $3.813$ and a semi-minor axis of $1.103$. The ratio of the semi-major axis to the radius is $1.472$; however, we have merely used half of the electric field. It shows that a relatively large electro-actuation can be achieved by using a relatively small electric field but suitably harnessing electromechanical instability. \\

%%%%%%%%%%%%%%%
%%%%%%%%%%%%%%%
%%%%%%%%%%%%%%%%%%%%%%%%%%%%%

%%%%%%%%%%%%%%%
\section{Concluding remarks} \label{sec-conclusion}
In conclusion, we have shown that either a pitchfork bifurcation (T-K instability) or a limit point bifurcation (pull-in instability), are achievable for a soft dielectric film. The applied electric field delays the onset of symmetry-breaking deformations and, more interesting, after the onset of T-K instability, an increased electric field can induce the symmetry-breaking state to revert back to the symmetric deformation prior to the onset of pull-in instability. The possibility of the rapid change of shapes between circular and elliptical films is capable of providing large actuation; in the conventional setting, the actuation is severely limited by the pull-in instability.\\

\section*{Conflicts of interest}
There are no conflicts to declare.\\

%%%%%%%%%%%%%%%
%%%%%%%%%%%%%%%

%%%%%%%%%%%%%%%%%%%%%%%%%%%%%%%%%%%%%%%%

\section*{Appendices}

%\appendix

\setcounter{equation}{0}
\renewcommand{\theequation}{A.\arabic{equation}}
\setcounter{figure}{0} \renewcommand{\thefigure}{A\arabic{figure}}
%%%%%%%%%%%%%%% 
%%%%%%%%%%%%%%% 
\section*{A. Details of the Formulation} \label{formulation}
Consider a circular dielectric film with radius $R$ and thickness $H$ in the undeformed state. Taking the Cartesian coordinates $(X_1, X_2, X_3)$ with an {\it orthonormal} basis $({\bm e}_1, {\bm e}_2, {\bm e}_3)$, the domain of the circular film in the reference configuration is represented by $\mathcal B_0 = \{\bm X \in {\mathbb R}^3: 0 \le \sqrt{X_1^2 + X_2^2} \le R, \ 0 \le X_3 \le H \}$. We consider the homogeneous thinning in which the deformation $\bm x$ has the following component form:
\begin{equation} \label{sy-6}
x_1 = \alpha_1 X_1, \quad x_2 = \alpha_2 X_2, \quad x_3 = \alpha_3 X_3,
\end{equation}
where $\alpha_i >0$, $i =1,2,3$, are constant stretches. By the deformation \eqref{sy-6}, the deformation gradient $\bm F =\nabla \bm x$ in the Cartesian coordinates is given by
\begin{equation} \label{sy-7}\emph{}
{\bm F} := {\rm diag} \, \left(\alpha_1, \alpha_2, \alpha_3 \right),
\end{equation}
which is represented by a diagonal matrix. For incompressible elastomers, the constraint of incompressibility requires that the Jacobian $J = \det {\bm F}$ must be one, i.e., 
\begin{equation} \label{sy-9}
\alpha_1 \alpha_2 \alpha_3 = 1.
\end{equation}
The total nominal stress ${\bm T}$ within the incompressible dielectric elastomers can be written as follows \cite{suo2008nonlinear, zhao2014harnessing}:\\
\begin{equation} \label{sy-10}
{\bm T} = \frac{\partial W^e}{\partial {\bm F}} + {\bm T}^M - \kappa {\bm F}^{-T}.
\end{equation}

Here $W^e$ is the strain-energy function of the purely elastic part, ${\bm T}^M$ is the nominal Maxwell stress, $\kappa$ serves as the Lagrange multiplier, and ${\bm F}^{-T}$ is the inverse of the transpose ${\bm F}^{T}$. For a linear dielectric film subjected to an applied voltage $\Phi$ in the thickness direction, the nominal Maxwell stress is given by ${\bm T}^M := \frac{1}{2} \varepsilon {\tilde E}^2 \alpha_3^{-2} {\rm diag} \,  (-\alpha_1^{-1}, -\alpha_2^{-1}, \alpha_3^{-1} )$, where $\varepsilon$ is the material permittivity and $\tilde E = \Phi/H$. \\

Assuming isotropy, the strain-energy function $W^e ({\bm F})$ in \eqref{sy-10} depends on the deformation gradient through the principal stretches $\alpha_1$, $\alpha_2$ and $\alpha_3$. Thus, the total nominal stress in \eqref{sy-10} is expressed as ${\bm T} := {\rm diag} \, (T_{1}, T_{2}, T_{3})$. The principal stresses are $T_{j} ={\partial W^e}/{\partial \alpha_j} - ({\varepsilon {\tilde E}^2 \alpha_3^{-2}}/{2} + \kappa)  \alpha_j^{-1}$, where $j=1,2$, and $T_{3}={\partial W^e}/{\partial \alpha_3} + ({\varepsilon {\tilde E}^2 \alpha_3^{-2}}/{2} - \kappa)  \alpha_3^{-1}$. With the boundary conditions $\bm T {\bm e}_3 = {\bm 0}$ on the top and bottom surfaces, we obtain the Lagrange multiplier
$\kappa = \alpha_3 {\partial W^e}/{\partial \alpha_3} + {\varepsilon {\tilde E}^2 \alpha_3^{-2}}/{2}$. Then the in-plane principal stresses are\\
\begin{equation} \label{new-T11-T22}
T_{j} = \frac{\partial W^e}{\partial \alpha_j} - \Big(  \alpha_3 \frac{\partial W^e}{\partial \alpha_3} + {\varepsilon {\tilde E}^2 \alpha_3^{-2}}  \Big)  \alpha_j^{-1},
\end{equation}\\
where $j=1,2$. Since the principal stresses are independent of the coordinates, the equilibrium equation, ${\rm Div} \, \bm T = \bm 0$ that are differential equations in general, are automatically satisfied here. \\

Finally, we only have to consider the traction boundary conditions at $\sqrt{X_1^2 + X_2^2}=R$, i.e., $\bm T {\bm e}_r = S {\bm e}_r$ with the normal ${\bm e}_r = \cos \theta {\bm e}_1 + \sin \theta {\bm e}_2$. Together with the diagonal matrix ${\bm T} := {\rm diag} \, (T_{1}, T_{2}, T_{3})$, we obtain
\begin{equation} \label{T11=S-T22=S-1}
S = T_{j},
\end{equation}
where $j=1,2$. The two algebraic equations \eqref{T11=S-T22=S-1}, together with the constraint of incompressibility \eqref{sy-9}, form a system of algebraic equations with unknowns $(\alpha_1, \alpha_2, \alpha_3)$ for a given pair of loads $(\tilde E, S)$. \\

In this paper, we consider a Mooney-Rivlin solid \cite{rivlin1948large} of which the strain-energy function in \eqref{new-T11-T22} is
\begin{equation} \label{mooney-rivlin-energy}
W^e = \frac{\mu}{2} \sum_{i=1}^3 \left\{ \left(\alpha_i^2 -1 \right) + \gamma \left(\alpha_i^{-2} -1 \right) \right\},
\end{equation}
where $\alpha_i$, $i =1,2,3$, are the principal stretches, $\mu$ and $\gamma$ are material constants. Typically, the case of $\gamma =0$ corresponds to the strain-energy function of a neo-Hookean solid. Both neo-Hookean and Mooney-Rivlin models are able to give good agreement with the experiment data at small and moderate strains. However, an apparent discrepancy is found at large strains. The Gent model is usually used to capture the stress-strain relation of an elastomer with nearly full stretched molecular chains \cite{gent1996new}. \\

It follows from \eqref{new-T11-T22}-\eqref{mooney-rivlin-energy} that\\
\begin{equation} \label{T11=S-T22=S-mooney}
\frac{S}{\mu}  = \left(\alpha_j - \gamma \alpha_j^{-3}\right) - \Big[ \alpha_3^2 + \Big( \frac{\varepsilon {\tilde E}^2}{\mu} - \gamma \Big) \alpha_3^{-2}  \Big] \alpha_j^{-1}.
\end{equation}\\

At a given pair of dead load and electric field, we seek the solution of $\alpha_1$, $\alpha_2$ and $\alpha_3$ from the two algebraic equations \eqref{T11=S-T22=S-mooney}, together with the constraint $\alpha_1 \alpha_2 \alpha_3 =1$. \\

Bearing in mind that the traction dead load $S$ is symmetric, and the applied electric field is homogeneous and is only applied in the thickness direction, a common assumption would be that the deformations of equilibrium states are symmetric, i.e., $\alpha_1 =\alpha_2$. However, we try to seek the possibility of symmetry-breaking deformations and discuss the ensuing electromechanical behavior. To directly show the relation between $\alpha_1$ and $\alpha_2$, we factor out $S$ and use $\alpha_1 \alpha_2 \alpha_3 = 1$ in \eqref{T11=S-T22=S-mooney}, then

\begin{equation} \label{T11=S-T22=S-mooney-1}
\begin{aligned}
0 &= (\alpha_1 - \alpha_2) \times \Big[1 + \Big( \frac{\varepsilon {\tilde E}^2}{\mu} -\gamma \Big) \alpha_1 \alpha_2 \\
& \quad + \alpha_1^{-3}\alpha_2^{-3} \left[ 1+  \gamma \left( \alpha_1^{2} + \alpha_1 \alpha_2 + \alpha_2^{2} \right) \right] \Big].
\end{aligned}
\end{equation}\\

Of interest is that the second term on the RHS of \eqref{T11=S-T22=S-mooney-1} is zero under some circumstances, which may allow the existence of equilibrium states with a symmetry-breaking deformation, i.e., $\alpha_1 \neq \alpha_2$. For neo-Hookean solids ($\gamma=0$), it is evident that the second term on RHS of \eqref{T11=S-T22=S-mooney-1} is always positive, then the identity \eqref{T11=S-T22=S-mooney-1} holds if and only if $\alpha_1 =\alpha_2$. In contrast, Mooney-Rivlin solids can give both symmetric and symmetry-breaking equilibrium solutions under some circumstances. The aforementioned statement highlights that too simple a constitutive choice (i.e. Neo-Hookean) may preclude observation of certain types of bifurcations.

\setcounter{equation}{0}
\renewcommand{\theequation}{B.\arabic{equation}}
\setcounter{figure}{0} \renewcommand{\thefigure}{B\arabic{figure}}

\section*{B. Details of the linear bifurcation analysis} \label{lba}
The linear bifurcation analysis here is actually the analysis of the uniqueness solutions $(\alpha_1, \alpha_2)$ to the two algebraic equations (6) in the main article. To proceed with the linear bifurcation analysis, we rewrite the two equations here:\\
\begin{equation} \label{bfa-b1}
L_j (\alpha_1, \alpha_2; \tilde E, S) = T_{j} (\alpha_1, \alpha_2; \tilde E)  - S = 0,
\end{equation}\\
where $j=1,2$. It should be noted that $\tilde E$ and $S$ are loading parameters while $\alpha_1$ and $\alpha_2$ are the unknown variables. For a given pair of loads $(\tilde E, S)$, the deformed film possesses the stretches $(\alpha_1, \alpha_2)$ through the solutions of the two algebraic equations \eqref{bfa-b1}. \\

By the implicit function theorem \cite{golubitsky2012singularities, chen2001singularity}, the necessary condition for the onset of bifurcation requires a zero determinant, namely
\begin{equation} \label{bfa-b2}
\left|
\begin{array}{cc}
\displaystyle\frac{\partial L_1}{\partial \alpha_1} & \displaystyle\frac{\partial L_1}{\partial \alpha_2}\\
\\
\displaystyle\frac{\partial L_2}{\partial \alpha_1} & \displaystyle\frac{\partial L_2}{\partial \alpha_2}
\end{array}
\right|
= 0.
\end{equation}
For Mooney-Rivlin solids, the explicit forms of equations \eqref{bfa-b1} are\\
\begin{subequations} \label{bfa-b3}
\begin{equation} \label{bfa-b3a}
L_1 = \alpha_1 - \gamma \alpha_1^{-3} - \alpha_1^{-3}\alpha_2^{-2} + (\gamma - \frac{\varepsilon {\tilde E}^2}{\mu}) \alpha_1 \alpha_2^{2} - \frac{S}{\mu} = 0, 
\end{equation}
\begin{equation} \label{bfa-b3b}
L_2 = \alpha_2 - \gamma \alpha_2^{-3} - \alpha_1^{-2}\alpha_2^{-3} +  (\gamma - \frac{\varepsilon {\tilde E}^2}{\mu}) \alpha_1^2 \alpha_2 - \frac{S}{\mu} = 0.
\end{equation}\\

To make a straightforward presentation of the relation between $\alpha_1$ and $\alpha_2$ in equilibrium, we subtract the two equations \eqref{bfa-b3a} and \eqref{bfa-b3b}, then we obtain \\
\begin{equation} \label{bfa-b3c}
\begin{aligned}
0 & =  (\alpha_1 - \alpha_2) \times \Big[1 + \Big( \frac{\varepsilon {\tilde E}^2}{\mu} -\gamma \Big) \alpha_1 \alpha_2  \\
& \quad + \alpha_1^{-3}\alpha_2^{-3} \left[ 1+  \gamma \left( \alpha_1^{2} + \alpha_1 \alpha_2 + \alpha_2^{2} \right) \right] \Big],
\end{aligned}
\end{equation}
\end{subequations}\\
which is exactly equation \eqref{T11=S-T22=S-mooney-1}. For the three equations \eqref{bfa-b3a}--\eqref{bfa-b3c}, any two can yield the solutions of $(\alpha_1, \alpha_2)$. The linear bifurcation analysis provides the conditions for the uniqueness solution. We now consider the entries of the matrix in \eqref{bfa-b2}. It follows from \eqref{bfa-b3a} and \eqref{bfa-b3b} that\\
\begin{equation} \label{bfa-b4}
\left\{
\begin{aligned}
\frac{\partial L_1}{\partial \alpha_1} & = 1 +3 \gamma \alpha_1^{-4} +3 \alpha_1^{-4}\alpha_2^{-2} + (\gamma - \frac{\varepsilon {\tilde E}^2}{\mu}) \alpha_2^{2}, \\
\frac{\partial L_1}{\partial \alpha_2} & = \frac{\partial L_2}{\partial \alpha_1} = 2 \alpha_1^{-3}\alpha_2^{-3} + 2 (\gamma - \frac{\varepsilon {\tilde E}^2}{\mu}) \alpha_1 \alpha_2, \\
\frac{\partial L_2}{\partial \alpha_2} & = 1 +3 \gamma \alpha_2^{-4} +3 \alpha_1^{-2}\alpha_2^{-4} +  (\gamma - \frac{\varepsilon {\tilde E}^2}{\mu}) \alpha_1^2.
\end{aligned}
\right.
\end{equation}\\

Consider a trivial solution that corresponds to the symmetric stretching $\alpha_1 = \alpha_2 = \alpha$. Equations \eqref{bfa-b3a} and \eqref{bfa-b3b} reduce to
\begin{equation} \label{bfa-b5}
\alpha - \gamma \alpha^{-3} - \alpha^{-5} + (\gamma - \frac{\varepsilon {\tilde E}^2}{\mu}) \alpha^{3} - \frac{S}{\mu} = 0,
\end{equation}
the entries \eqref{bfa-b4} become
\begin{equation} \label{bfa-b6}
\left\{
\begin{aligned}
\frac{\partial L_1}{\partial \alpha_1} & = \frac{\partial L_2}{\partial \alpha_2} = 1 +3 \gamma \alpha^{-4} +3 \alpha^{-6} + (\gamma - \frac{\varepsilon {\tilde E}^2}{\mu}) \alpha^{2}, \\
\frac{\partial L_1}{\partial \alpha_2} & = \frac{\partial L_2}{\partial \alpha_1} = 2 \alpha^{-6} + 2 (\gamma - \frac{\varepsilon {\tilde E}^2}{\mu}) \alpha^2,
\end{aligned}
\right.
\end{equation}
and the condition \eqref{bfa-b2} gives 
\begin{equation} \label{bfa-b7}
\begin{aligned}
0 & = \left( 1 +3 \gamma \alpha^{-4} +5 \alpha^{-6} + 3(\gamma - \frac{\varepsilon {\tilde E}^2}{\mu}) \alpha^{2} \right) \\
& \quad \times \left( 1 +3 \gamma \alpha^{-4} + \alpha^{-6} - (\gamma - \frac{\varepsilon {\tilde E}^2}{\mu}) \alpha^{2} \right).
\end{aligned}
\end{equation}

\setcounter{equation}{0}
\renewcommand{\theequation}{C.\arabic{equation}}
\setcounter{figure}{0} \renewcommand{\thefigure}{C\arabic{figure}}

\section*{C. Neo-Hookean vs Mooney-Rivlin Constitutive Law} \label{appendix-12loading}
\begin{figure}[ht!] % 
\centering
\centering
\subfigure[]{%
\includegraphics[width=0.85 \linewidth]{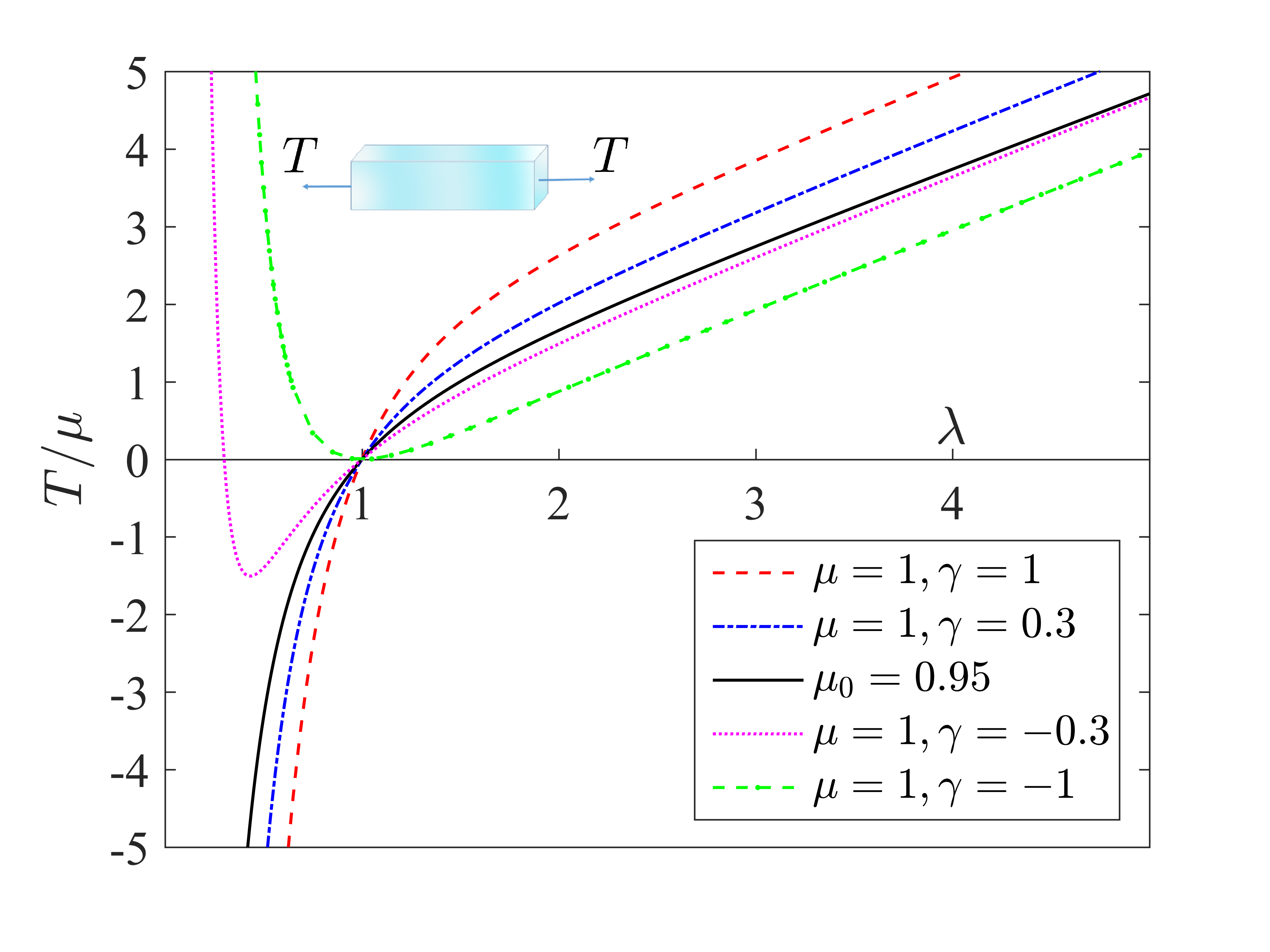} \label{afig1a}}
\subfigure[]{%
\includegraphics[width=0.85 \linewidth]{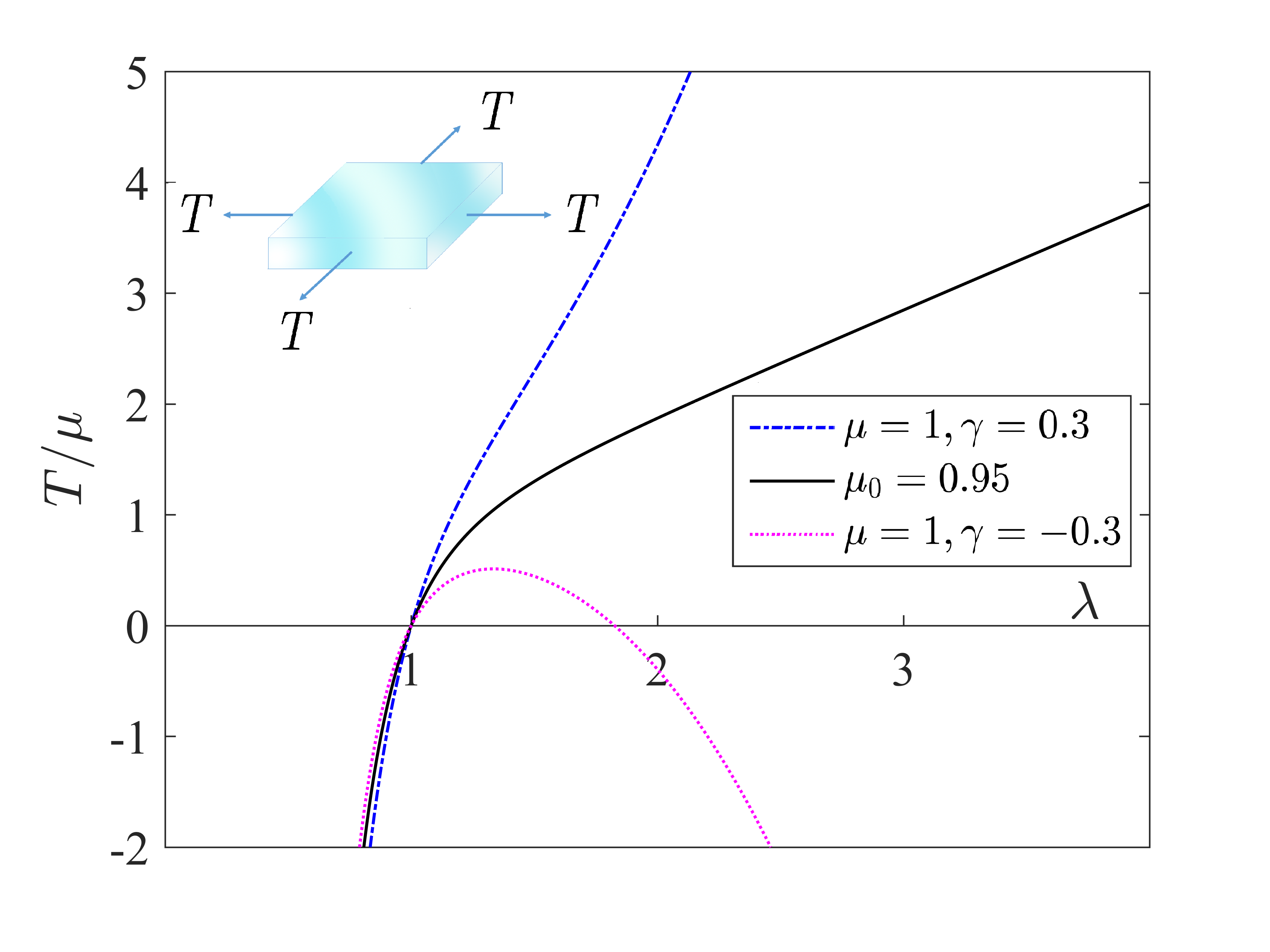} \label{afig1b}}
\caption{\small Loading of neo-Hookean and Mooney-Rivlin solids with different material parameters. (a) Stress-stretch curve for uniaxial loading. For neo-Hookean solids with a shear modulus of $\mu_0 = 0.95 \mu$, the stress-stretch curve can also be represented by a Mooney-Rivlin solid with material parameters $(\mu, \gamma) = (1, 0.3)$ qualitatively and quantitatively. (b) Stress-stretch curve for equibiaxial loading. Note that for negative parameters $\gamma<0$, a tensile stress $(T>0)$ in (a) can even lead to compression $(\lambda<1)$; moreover, a compressive stress $(T<0)$ in (b) can make a tension $(\lambda>1)$. To exclude these unusual stress-stretch responses, we thus omit the discussion of the case of $\gamma<0$ in the electromechanical instabilities in the main article.}
\label{afig1}
\end{figure}

In this section, we briefly contrast two types of hyperelastic constitutive laws--the Neo-Hookean vs Mooney-Rivlin by examining the loading of a cube (see Figure \ref{afig1}). Specifically, We show how  the material parameters affect the stress-strain curve, especially the unusual stress-strain response in Mooney-Rivlin solids with a negative parameter ($\gamma<0$). That is the reason  we drop the discussion of the case of $\gamma<0$ in the main article.

%%%%%%%%%%%%%%%
\subsection*{C.1 Uniaxial loading} \label{appendix-uniaxial}
Subjected to a uniaxial loading in the $X_1$ direction, the cube deforms from $1$ to $\lambda$ in the $X_1$ direction. Due to the constraint of incompressibility, the cube deforms from $1$ to $\lambda^{-1/2}$ in the $X_2$ and $X_3$ directions, respectively. Consider the neo-Hookean model of hyperelastic materials. The strain-energy function of the cube can be expressed as
\begin{equation} 
W^e = \frac{\mu_0}{2} \left(\lambda^2 + 2 \lambda^{-1} -3 \right),
\end{equation}
where $\mu_0$ is the shear modulus at small deformation. The nominal stress $T$ in the $X_1$ direction is
\begin{equation} 
T = \frac{\partial W^e}{\partial \lambda} = \mu_0 (\lambda - \lambda^{-2}).
\end{equation}

Consider the Mooney-Rivlin model that makes the strain-energy function of the cube as
\begin{equation} 
W^e = \frac{\mu}{2} \left\{ \left(\lambda^2 + 2\lambda^{-1} -3 \right) + \gamma \left(\lambda^{-2} + 2\lambda -3 \right) \right\},
\end{equation}
where $\mu$ and $\gamma$ are material constants. The nominal stress $T$ in the $X_1$ direction becomes
\begin{equation} 
T = \frac{\partial W^e}{\partial \lambda} = \mu \left\{(\lambda - \lambda^{-2}) + \gamma \left(- \lambda^{-3} + 1 \right) \right\}.
\end{equation}

%%%%%%%%%%%%%%%
\subsection*{C.2 Equibiaxial loading} \label{appendix-biaxial}
Subjected to the equibiaxial loading, the cube deforms from $1$ to $\lambda$ in both the $X_1$ and $X_2$ directions. Due to the constraint, the cube deforms from $1$ to $\lambda^{-2}$ in the $X_3$ directions. In this example, we exclude the discussion of asymmetric deformation. Consider the neo-Hookean model of hyperelastic materials. Then the strain-energy function of the cube is expressed as
\begin{equation}
W^e = \frac{\mu_0}{2} \left(2\lambda^2 + \lambda^{-4} -3 \right),
\end{equation}
and the equibiaxial nominal stress $T$ in the $X_1$ and $X_2$ directions is
\begin{equation} \label{neo-hookean-stress} 
T = \frac{1}{2} \frac{\partial W^e}{\partial \lambda} = \mu_0 (\lambda - \lambda^{-5}).
\end{equation}

If we take the Mooney-Rivlin model, the strain-energy function of the cube is 
\begin{equation} 
W^e = \frac{\mu}{2} \left\{ \left(2\lambda^2 + \lambda^{-4} -3 \right) + \gamma \left(2\lambda^{-2} + \lambda^4 -3 \right) \right\},
\end{equation}
and the nominal stress $T$ in the $X_1$ and $X_2$ directions is
\begin{equation} \label{mooney-rivlin-stress}
T = \frac{1}{2}\frac{\partial W^e}{\partial \lambda} = \mu \left\{(\lambda - \lambda^{-5}) + \gamma \left(- \lambda^{-3} + \lambda^3 \right) \right\}.
\end{equation}

%%%%%%%%%%%%%%%%%%%%%%%%%%%%%

\setcounter{equation}{0}
\renewcommand{\theequation}{D.\arabic{equation}}
\setcounter{figure}{0} \renewcommand{\thefigure}{D\arabic{figure}}

%%%%%%%%%%%%%%%%%
%%%%%%%%%%%%%%%%%Appendix
\section*{D. Free energy of the electrostatic system} \label{appendix-energy-method}
The linear bifurcation analysis can neither distinguish the type of bifurcation nor determine the stability of the bifurcated branch. We therefore carry out the stability analysis by using the energy method. Consider the free energy of an electrostatic system \cite{suo2008nonlinear, zhao2007method}. Then the free energy $G$ of a circular dielectric film subjected to an applied voltage $\Phi$ in the thickness direction and an in-plane symmetric dead load $S$ can be written as
\begin{equation} 
G =\pi R^2 H  W(\alpha_1, \alpha_2, \tilde D) - \int_{\partial \mathcal B_0} S {\bm e}_r \cdot (\bm x - \bm X) - \Phi Q,
\end{equation}
where ${\partial \mathcal B_0}$ is the lateral surface, i.e., ${\partial \mathcal B_0} = \{\bm X \in {\mathbb R}^3: \sqrt{X_1^2 + X_2^2} = R, \ 0 \le X_3 \le H \}$. Consider the
homogeneous thinning (1) in the main article, i.e., $x_1 = \alpha_1 X_1, x_2 = \alpha_2 X_2, x_3 = \alpha_3 X_3$. On the boundary ${\partial \mathcal B_0}$, we have the displacement
\begin{equation} 
\bm x - \bm X = (\alpha_1 -1) R \cos \theta {\bm e}_1 + (\alpha_2 -1) R \sin \theta {\bm e}_2,
\end{equation}
and the inner product is
\begin{equation} 
S {\bm e}_r \cdot (\bm x - \bm X) = S R [ \alpha_1 \cos^2 \theta + \alpha_2 \sin^2 \theta -1],
\end{equation}
where $0 \le \theta \le 2 \pi$. Thus,
\begin{equation} 
\begin{aligned}
& \int_{\partial \mathcal B_0} S {\bm e}_r \cdot (\bm x - \bm X) \\
& = \int_0^{2\pi} S R^2 H [ \alpha_1 \cos^2 \theta + \alpha_2 \sin^2 \theta -1] d \theta \\
& = S \pi R^2 H (\alpha_1 + \alpha_2 -2).
\end{aligned}
\end{equation}

On the other hand, we consider the following form
\begin{equation}
\Phi Q = \pi R^2 H \tilde E \tilde D,
\end{equation}
where the nominal electric field is $\tilde E =\Phi/H$ and the nominal electric displacement is $\tilde D = Q/(\pi R^2)$. Now the free energy density is
\begin{equation} \label{S21}
g = \frac{G}{\pi R^2 H} = W (\alpha_1, \alpha_2, \tilde D) - S (\alpha_1 + \alpha_2 -2) - \tilde E \tilde D.
\end{equation}

For linear dielectrics, the energy function of the Mooney-Rivlin type dielectrics is
\begin{equation} \label{S22}
\begin{aligned}
 W (\alpha_1, \alpha_2, \tilde D) & = \frac{\mu}{2} \left(\alpha_1^{2} + \alpha_2^{2} + \alpha_1^{-2} \alpha_2^{-2} -3 \right)    \\
& \quad +  \frac{\mu}{2}  \gamma \left(\alpha_1^{-2} + \alpha_2^{-2} + \alpha_1^{2}\alpha_2^{2} -3 \right) \\
& \quad + \frac{{\tilde D}^2}{2 \varepsilon} \alpha_1^{-2} \alpha_2^{-2}.
\end{aligned}
\end{equation}
By substituting \eqref{S22} into \eqref{S21}, the equilibrium in electric qualities, ${\partial g}/{\partial \tilde D} =0$, gives
\begin{equation} \label{S23}
\tilde E = \frac{{\tilde D}}{\varepsilon} \alpha_1^{-2} \alpha_2^{-2}.
\end{equation}
It follows from \eqref{S22} and \eqref{S23} that the free energy density $g$ in \eqref{S21} becomes
\begin{equation} 
\begin{aligned}
g & = \frac{\mu}{2} \left(\alpha_1^{2} + \alpha_2^{2} + \alpha_1^{-2} \alpha_2^{-2} -3 \right) \\
& \quad + \frac{\mu}{2} \gamma \left(\alpha_1^{-2} + \alpha_2^{-2} + \alpha_1^{2}\alpha_2^{2} -3 \right) \\
& \quad - S (\alpha_1 + \alpha_2 -2) - \frac{\varepsilon{\tilde E}^2}{2} \alpha_1^{2} \alpha_2^{2}.
\end{aligned}
\end{equation}

Consider the normalizations
\begin{equation}
\hat g = \frac{g}{\mu}, \quad \hat S = \frac{S}{\mu}, \quad \hat E = \frac{\tilde E}{\sqrt{\mu/\varepsilon}}.
\end{equation}

We then have the normalized free energy density
\begin{equation} 
\begin{aligned}
\hat g & = \frac{1}{2} \left(\alpha_1^{2} + \alpha_2^{2} + \alpha_1^{-2} \alpha_2^{-2} -3 \right) \\
& \quad + \frac{1}{2} \gamma \left(\alpha_1^{-2} + \alpha_2^{-2} + \alpha_1^{2}\alpha_2^{2} -3 \right) \\
& \quad - \hat S (\alpha_1 + \alpha_2 -2) - \frac{{\hat E}^2}{2} \alpha_1^{2} \alpha_2^{2}.
\end{aligned}
\end{equation}\\

For certain loads $\hat S$ and $\hat E$ with a given material parameter $\gamma$, we can plot the contour of the free energy density $\hat g$ on the $\alpha_1 - \alpha_2$ plane.\\

%\vspace{0.5 in} 

{\bf{Acknowledgements}}---P.S. acknowledges support from the University of Houston M.D. Anderson Professorship. S.Y. acknowledges Qilu Youth Scholar Program of Shandong University, the Fundamental Research Funds for the Central Universities (2020JCG012), Natural Science Foundation of Jiangsu Province (SBK2020041645) and the Program of Science and Technology of Suzhou (SYG202005). B.W. is supported by the National Key Research and Development Program of China (2018YFB0703500). K. D. acknowledges support from NSF (1635407), ARO (W911NF-17-1-0084), ONR (N00014-18-1-2528, N00014-18-1-2856), BSF (2018183), and AFOSR (MURI FA9550-18-1-0095). \\

{
\small{
	\bibliographystyle{ieeetr}
	\bibliography{reference}
}
}

\end{document}